\documentstyle[preprint,eqsecnum,aps]{revtex}

\tightenlines
\begin{document}
\draft
\title{\bf Finite-Size Scaling in the transverse Ising Model on a Square
Lattice 
} 
\author{C. J. Hamer}  
\address{School of Physics,    \\                                           
The University of New South Wales,        \\                            
Sydney, NSW 2052, Australia.}                      

%\date{September 20, 1995}
\maketitle 
\begin{abstract}
Energy eigenvalues and order parameters are calculated by exact
diagonalization for the transverse Ising model on square lattices
of up to 6x6 sites. 
Finite-size scaling is used to estimate the critical parameters of the model, confirming universality with the
three-dimensional classical Ising model. Critical amplitudes are also estimated for both the energy gap and the
ground-state energy.
\end{abstract}
\pacs{PACS Indices: 02.70.+d,05.30.-d,05.50.+q,64.40.Cn \\ 
\\  \\ 
(Submitted to  J. Phys. A) }
\newpage

%\narrowtext

\section{Introduction}

      Recent advances in computer technology have allowed the exact
diagonalization of Ising-type quantum spin systems up to 36 sites in
size. Schulz, Ziman and Poilblanc (1996), for example, studied the J1-J2 XXZ
Heisenberg spin model on square lattices up to 6x6 sites. 
Our aim in this paper is to carry out an exact diagonalization study of
the transverse Ising model on the square lattice, in order to estimate
its critical parameters and study its finite-size scaling behaviour.

	The transverse Ising model in (2+1)D is well-known to be the
quantum Hamiltonian corresponding to the classical 3D Ising model
(Suzuki 1976, Fradkin and Susskind 1978), and
exhibits a quantum phase transition in the same universality class as
the classical 3D Ising thermal transition. 
It was first studied by series expansion methods by Pfeuty and Elliott
(1971), and there have been several further series expansion
calculations since then, both `low-temperature' (Marland 1981, Yanase et
al 1976, Oitmaa et
al 1991) 
and `high-temperature' (Hamer and Irving 1984, Hamer and Guttmann
1989, He et al 1990).
Exact finite-lattice calculations have also been carried out previously
(Roomany and Wyld 1980, Hamer 1983, Henkel 1984, 1987) for square lattices of up to 5x5
sites, and similar calculations have also been done for the triangular
lattice (Hamer and Johnson 1986, Henkel 1990, Price et al 1993).
Here we extend these calculations for the first time to the 6x6 lattice,
and use finite-size scaling theory to obtain improved estimates of the
critical point and critical index $\nu$.

	The finite-size scaling amplitudes at the critical point are
also of interest. In (1+1)D, it is well-known that the theory of
conformal invariance relates the scaling amplitudes to fundamental
parameters of the underlying effective field theory at the critical
point, such as the conformal anomaly and scaling indices. In higher
dimensions, a similar scenario is known to hold at `first-order'
transitions, where a continuous symmetry is spontaneously broken, giving
rise to Goldstone bosons (Hasenfratz and Leutwyler 1990): the finite-size
scaling amplitudes are related to parameters of the Goldstone bosons
such as the spin-wave stiffness and spin-wave velocity. Does something
similar apply at second-order transitions in higher dimensions? Apart
from a discussion by Cardy (1985), little has been done in this area. One
peculiar result was obtained by Henkel (1986, 1987) and Weston (1990), and confirmed recently by
Weigel and Janke (1999): the scaling amplitudes of the spin-spin and
energy-energy correlation lengths on {\it antiperiodic} lattices have a
universal ratio:
\begin{equation}
\frac{A_{\sigma}}{A_{\epsilon}} = \frac{x_{\sigma}}{x_{\epsilon}}
\end{equation}
where the $x_{i}$ are the scaling indices in the respective sectors.
This phenomenon appears to have no good theoretical explanation at the
present time.

	Our exact diagonalization methods are outlined briefly in
Section 2, and the numerical results are presented in Section 3. The
critical parameters so obtained are compared with other estimates in
Section 4, and the critical amplitudes are discussed.

\section{Method}

      The transverse Ising model on the square lattice has 
the Hamiltonian
\begin{eqnarray}
 H = \sum_{i}(1 - \sigma_{3}(i)) 
   - x\sum_{<ij>}\sigma_{1}(i)\sigma_{1}(j)
   - h\sum_{i}\sigma_{1}(i)
\label{2.1}
\end{eqnarray}
where the sum $<ij>$ runs over nearest neighbour pairs on the lattice,
and the $\sigma$ matrices are the usual Pauli spin operators acting on a
2-state spin-variable at each site. 
The coupling x is analogous to an inverse `temperature', and h
represents an external `magnetic field'.
We shall employ a representation
in which the {$\sigma_{3}(i)$} are diagonal. 
Periodic boundary conditions are assumed.

	The unperturbed ground-state of the model at $x = 0$ has all spins `up', i.e. $\sigma_{3}(i) =
+1$, all $i$. The interaction term will induce an admixture of states with `flipped' spins. The Hilbert
space of the model consists of two sectors, containing an odd and even number of flipped spins
respectively.

      Exact diagonalizations have been carried out for LxL lattices, L =
1,..,6. The methods employed are fairly standard, for the most part, and
will not be described in detail here. First, a list of allowed basis
states in the given sector was prepared, using the `sub-lattice coding'
technique of Lin (1990). This efficient technique produces a sorted list
of states, requiring only one integer word of storage per state. Since
only the zero-momentum states are considered here, the states were
`symmetrized': that is, all copies of a given state under translations,
reflections and rotations were represented by a single state. Thus for
the 6x6 lattice in the even sector, the total number of
`unsymmetrized' states is approximately $2^{35}$, whereas under
symmetrization this is reduced by a factor of approximately 288, down to
119,539,680.

      Next, the Hamiltonian matrix elements are generated, by applying
the interaction operators of equation (\ref{2.1}) to each initial state,
symmetrizing the resulting final state, and looking it up in the master
file. The elements were grouped into blocks, each of which acts between
small sub-sets of the initial and final state vectors, to avoid
`thrashing' during the matrix multiplications. Within each sub-set, the
initial and final addresses can each be fitted into a half-integer, so that
the matrix elements occupied 35 Gbyte of storage over all.

      Finally, the lowest eigenvalue and eigenvector of the Hamiltonian
were found in each sector, using the conjugate gradient method. Nightingale et al (1993)
showed that the conjugate gradient method converges faster than the
Lanczos method for large problems such as this. We found that the eigenvalue converged to an accuracy of
1 part in $10^{10}$ in 20-25 iterations for the 6x6 lattice in the neighbourhood of the critical
point. 

	Having determined the quantities of interest for each finite
lattice, it is then necessary to make an extrapolation to $L \rightarrow
\infty$, to estimate the bulk behaviour of the system. In the vicinity
of the critical point, the finite-lattice sequence will typically behave
as
\begin{equation}
f_{L} = f_{\infty} + a_{1}L^{-\omega_{1}} + a_{2}L^{-\omega_{2}} +
\cdots
\label{2.2}
\end{equation}
where the $\omega_{i}$ are non-integer exponents, in general (Barber 1983). The
problem of extrapolating such a sequence has been discussed in several
reviews (Smith and Ford 1982, Barber and Hamer 1982, Guttmann 1989). We have employed a number of different algorithms,
including: 
\begin{itemize}
\item[i)] the Neville table (Guttmann 1989), which is best suited to a simple
polynomial sequence, with integer exponents $\omega_{i}$;
\item[ii)] the alternating VBS algorithm (van den Broeck and Schwartz 1979, Barber and Hamer
1982), which can give good
convergence for sequences of type (\ref{2.2}), but needs at least two iterations to
work well;
\item[iii)] the Lubkin algorithm (1952), which is more suitable for
short sequences;
\item[iv)] the Bulirsch-Stoer algorithm (1964), which has been applied
in this context by Henkel and Patkos (1987), and Henkel and Sch{\"
u}tz (1988). This algorithm involves an explicit parameter $\omega$
which can be optimized to match the leading power-law correction. It has
been claimed by Henkel and Sch{\" u}tz that the algorithm is more robust
and more accurate than the VBS algorithm, especially for short
sequences.

\end{itemize}

\section{Results}

\subsection{Finite-lattice data}

	The pseudo-critical point at lattice size L can be defined according to finite-size scaling
theory (Barber 1983) as the coupling $x_{L}$ such that 
\begin{equation}
R_{L}(x_{L}) = 1
\end{equation}
where $R_{L}(x)$ is the scaled energy-gap ratio
\begin{equation}
R_{L}(x) = \frac{LF_{L}(x)}{(L-1)F_{L-1}(x)} 
\end{equation}
and $F_{L}(x)$ is the energy gap for lattice size L. This point is found by calculating the energy
eigenvalues at a cluster of 5 equally spaced points in the neighbourhood of $x_{L}$, and then finding
$x_{L}$ by interpolation between them. The spacing between the points was chosen as $\Delta x =
0.001$, estimated to balance the truncation and round-off errors in the calculation. The values of all
other observables can then be estimated at $x_{L}$ by the same finite-difference interpolation procedures.
	Tables 1 and 2 list the pseudo-critical points $x_{L}$, and the values of the calculated observables
at coupling $x_{L}$ for each pair of lattice sizes L and (L-1), for  L = 2,3,4,5 and 6. 
The values of $x_{L}$ for $L = 2$ to 5 listed in Table 1 agree through
six figures with those calculated previously (Hamer 1983).

	Table 1 lists values for the ground-state energy per site
$\epsilon_{0,L}$ for lattice size L, and its derivatives
$\epsilon'_{0,L}$ and $\epsilon''_{o,L}$, where the prime denotes
differentiation with respect to x. The values are expected to be
accurate to the figures quoted (or better) as regards round-off error.
The truncation error in the 5-point interpolation process is harder to
estimate, since it involves unknown higher derivatives of
$\epsilon_{0}$, but we estimate it should be no more than about 1 part
in $10^{12}$ for $\epsilon_{0}$ and 1 part in $10^{6}$ for $\epsilon_{0}''$.

	We have also listed values in Table 1 for the magnetic
susceptibility, defined by 
\begin{equation}
\chi_{L} = -\frac{1}{L^{2}}\frac{\partial^{2}E_{0,L}(x,h)}{\partial h^{2}} |_{h =0}
\end{equation}
This derivative was also estimated by a finite difference method, using
a cluster of 5 data points around $h = 0$, with a spacing $\Delta h =
0.0003$, giving an estimated truncation error of no more than 1 part in $10^{6}$ in
the susceptibility. This calculation was a little too large to carry
through for $L = 6$, with the facilities available.

	Table 2 lists the energy gap $F_{L}$ between the odd and even
sectors, and its derivatives $F'_{L}$ and $F''_{L}$, at each $x_{L}$.
Values are also listed here for the quantity $M_{L}$ defined by
\begin{equation}
M_{L} = \langle 0 | \sigma_{1}(1) | 1 \rangle
\label{3.4}
\end{equation}
where $| 0 \rangle$, $| 1 \rangle$ are the lowest-lying energy
eigenvectors in the even and odd sectors, respectively. It can be shown (Yang 1952, Uzelac 1980, Hamer 1982) that this quantity converges towards the
spontaneous magnetization in the bulk limit. Unfortunately, for
technical reasons we were again unable to calculate this quantity for
L=6. Since the accuracy of the wavefunction is only the square root of
that of the eigenvalue, the round-off error in these values is expected to
be about 1 part in $10^{6}$.

\subsection{Critical Point}

	The sequence of pseudo-critical points $x_{L}$ converges rapidly, as can be seen in Fig. 1, where
$x_{L}$ is plotted against $1/L^{4}$.
To estimate the bulk limit ($L \rightarrow \infty$) of this sequence, we
have employed various algorithms discussed above,
as well as a simple polynomial fit in $1/L^{4}$ and higher powers. 

Our final estimate of the critical point is
\begin{equation}
x_{c} = 0.32841(2)
\label{3.5}
\end{equation}
This is consistent with our earlier finite-size estimate of $x_{c} = 0.3289(10)$
 (Hamer 1983), but nearly two orders of magnitude more
accurate. Henkel (1987) obtained an improved estimate $x_{c} = 0.3282(1)$ from lattices up to 5x5
sites.

\subsection{Critical Indices}

	Finite-size scaling theory (Barber 1983) also tells us how to estimate the critical indices for the model.
The finite-lattice susceptibility $\chi_{L}$, for instance, is predicted
to scale at the critical point like
\begin{equation}
\chi_{L}(x_{c}) \sim L^{\gamma/\nu}, \hspace{5mm} L \rightarrow \infty
\label{3.6}
\end{equation}
and hence one finds that 
\begin{equation}
L(1 - \frac{\chi_{L}(x_{L})}{\chi_{L-1}(x_{L})}) \sim -
\frac{\gamma}{\nu}, \hspace{5mm} L \rightarrow \infty
\label{3.7}
\end{equation}
Similarly, ratios of the finite-lattice `magnetizations' (equation
\ref{3.4}) give estimates of
$\beta/\nu$. 
Finally,
estimates of the index $1/\nu$ 
can be obtained from the Callan-Symanzik `beta function' (Barber 1983),
\begin{equation}
\beta_{L}(x)/g = \frac{F_{L}(x)}{(F_{L}(x)-2xF'_{L}(x))}
\label{3.9}
\end{equation}
via
\begin{equation}
 L (1-\frac{\beta_{L}(x_{L})}{\beta_{L-1}(x_{L})})
 \sim \frac{1}{\nu}, \hspace{5mm} L \rightarrow \infty
\label{3.10}
\end{equation}

	One would expect to obtain estimates of the ratio $\alpha/\nu$
in a similar fashion from the `specific heat',
\begin{equation}
C_{L}(x) = -\frac{x^{2}}{L^{2}}\frac{\partial^{2}\epsilon_{0}}{\partial
x^{2}},
\label{3.8}
\end{equation}
but it is known (Hamer 1983) that these estimates are very poor, too
high by a factor of nearly 2. The reason is easily found: the
ground-state energy or specific heat contains a `regular' or analytic
piece as well as the singular term (Privman and Fisher 1984). Henkel (1987) has
cleverly sidestepped this problem, using a transition amplitude to find
$\alpha/\nu$, in analogy to equation (\ref{3.4}). Here, we eliminate the
regular term by subtracting:
\begin{equation}
\epsilon_{0,L}'' - \epsilon_{0,L-1}'' \sim L^{\alpha/\nu - 1},
\hspace{5mm} L \rightarrow \infty,
\label{3.10a}
\end{equation}
and using successive ratios of these differences to estimate
$1 -\alpha/\nu$
The estimates so obtained for the critical index ratios are listed in
Table 3.

	Alternatively, `logarithmic' estimates of the critical indices
may be obtained as follows:
\begin{equation}
\frac{\ln[\chi_{L}(x_{L})/\chi_{L-1}(x_{L})]}{\ln[L/(L-1)]} \sim
\frac{\gamma}{\nu}, \hspace{5mm} L \rightarrow \infty
\label{3.11}
\end{equation}
These alternative estimates are listed in Table 4.
The finite-size corrections are generally smaller for the logarithmic
estimates.

	These finite-size estimates of the critical indices agree
closely with the previous calculation of Hamer (1983) up to $L =5$; and
remarkably enough, most of them agree to within 4 significant figures
with the equivalent results obtained for the triangular lattice (Hamer
and Johnson 1986, Price et al 1993).

	The same algorithms mentioned above have been employed to
extrapolate these sequences to their bulk limit. 
The sequences are very short, and may have slight irregularities, so
that the tabular algorithms are generally no more accurate than simple
graphical methods or polynomial fits in the extrapolation.
The resulting estimates
are listed at the foot of Tables 3 and 4. The errors in these estimates
are inevitably rather subjective, but the variation between different
algorithms gives some indication of the likely error.

	Figure 2 graphs the estimates of $1/\nu$ from Tables 3 and 4 
as a function of
$1/L$: it can be seen that the behaviour is almost precisely linear for
the estimates from Table 3. 
Correspondingly, the Neville tables and polynomial fits give stable
results, while the Lubkin and Bulirsch-Stoer algorithms give less stable
results, possibly a little higher. We conclude that
\begin{equation}
\frac{1}{\nu} = 1.591(2) 
\end{equation}

The estimates for $\alpha/\nu$ are not quite so well-behaved, but our
final estimate is
\begin{equation}
\frac{\alpha}{\nu} = 0.16(1) 
\end{equation}
This is a much better result than can be obtained directly from the
specific heat, equation (\ref{3.8}).

	The estimates for the other indices $\beta/\nu$ and $\gamma/\nu$
are rapidly convergent, but we only have data up to $L = 5$, which were
known previously. We find
\begin{equation}
\frac{\beta}{\nu} = 0.522(2) 
\end{equation}
\begin{equation}
\frac{\gamma}{\nu} = 1.96(1) 
\end{equation}

\subsection{ Energy Amplitudes}

	The finite-size behaviour of the energy gap at the critical
point is 
\begin{equation}
F_{L}(x_{L}) \sim \frac{A_{1}}{L}, \hspace{5mm} L \rightarrow \infty
\end{equation}
so the amplitude $A_{1}$ can be estimated by
\begin{equation}
LF_{L}(x_{L}) \sim A_{1}, \hspace{5mm} L \rightarrow \infty
\end{equation}
The sequence of estimates for $A_{1}$ is shown in Figure 3. It
extrapolates to a value
\begin{equation}
A_{1} = 1.39(1)
\end{equation}
A value of 1.42 was previously estimated by Henkel (1987).

The finite-size scaling behaviour of the ground-state energy per site
$\epsilon_{0}$ at the pseudo-critical point is shown in Figure 4. The
finite-size scaling corrections appear to decrease like $1/L^{3}$, 
in accordance with the Privman-Fisher scaling hypothesis (1984), which
states that the singular part of the free energy density of a system of
finite size L should scale as $L^{-d}$ (here d = 3). A
polynomial fit on this assumption gives
\begin{equation}
\epsilon_{0,L}(x_{L}) \sim \epsilon_{0}^{*} -
\frac{A_{0}}{L^{3}}, \hspace{5mm} L \rightarrow \infty
\end{equation}
with
\begin{equation}
\epsilon_{0}^{*} = -0.624(1) 
\end{equation}
and
\begin{equation}
A_{0} = 0.38(5)
\label{3.21}
\end{equation}
Further evidence for this power-law behaviour can be obtained as
follows. Suppose
\begin{equation}
\epsilon_{0,L}(x_{L}) \sim \epsilon_{0}^{*} -
A_{0}/L^{p}, \hspace{5mm} L \rightarrow \infty
\end{equation}
then 
\begin{equation}
L[1-\frac{(\epsilon_{0,L}(x_{L})-\epsilon_{0,L-1}(x_{L}))}
{(\epsilon_{0,L-1}(x_{L-1})-\epsilon_{0,L-2}(x_{L-1}))}] \equiv p_{L} \sim
p \hspace{5mm} $as $ L \rightarrow \infty
\end{equation}
and
\begin{equation}
\ln[\frac{\epsilon_{0,L}(x_{L})-\epsilon_{0,L-1}(x_{L})}
{\epsilon_{0,L-1}(x_{L-1})-\epsilon_{0,L-2}(x_{L-1})}]/\ln[L/L-1] \sim
-p 
 \hspace{5mm} $as $ L \rightarrow \infty
\end{equation}
The sequences of finite lattice estimates for p are shown in Figure 5. It
can be seen that the `linear' sequence comes down towards 3 from above,
whilst the `logarithmic' sequence comes up towards 3 from below. The
sequences are a little irregular, however, and the best estimate we can
obtain for the bulk limit is
\begin{equation}
p = 2.8(2)
\end{equation}
a little lower than, but still consistent with 3.

	Assuming that $p = 3$, the scaling amplitude $A_{0}$ for the
ground-state energy can be found by
\begin{equation}
\frac{L^{4}}{3}(\epsilon_{0,L}(x_{L})-\epsilon_{0,L-1}(x_{L})) \sim A_{0},
\hspace{5mm} L \rightarrow \infty
\end{equation}
The sequence of estimates for $A_{0}$ is graphed in Figure 6, and
extrapolates to a value
\begin{equation}
A_{0} = 0.35(2),
\end{equation}
which is in reasonable agreement with equation (\ref{3.21}). 
Henkel (1987) previously obtained an estimate of 0.39 for this quantity
(allowing for the different normalization of his Hamiltonian).
Mon (1985) obtained a Monte Carlo estimate of the corresponding free
energy amplitude in the 3D classical model.
	
	In order to `calibrate' this result, we need to know the ``speed
of light' v, or in other words the scale factor needed in this model to
make the long-range correlations isotropic in space and time at the
critical point. We have attempted to estimate this 
using the dispersion relation for the lowest excited state at the
critical point, expected to be of the form
\begin{equation}
E(k) = vk
\end{equation}
in the bulk system. We have calculated the finite-lattice eigenvalues
for low-lying excited states with non-zero momntum for lattice sizes L=2
to 5, and set
\begin{equation}
v_{L} = \frac{L}{2\pi}(F_{L}(x_{L},\frac{2\pi}{L})-F_{L}(x_{L},0))
\sim v, \hspace{5mm}L \rightarrow \infty
\end{equation}
where $F_{L}(x,k)$ is the energy at coupling x for momentum k.
Figure 7 shows the sequence of finite-lattice estimates for v as a
function of $1/L$. They extrapolate to a bulk value
\begin{equation}
v = 0.99(3) 
\end{equation}

	The ratio $A_{0}/v$ should be a universal number, independent of
the normalization of the Hamiltonian. From the results above, we find
\begin{equation}
\frac{A_{0}}{v} = 0.35(2),
\label{3.31}
\end{equation}
to be compared with values of 0.719 expected according to effective
field theory for a single free boson (Hasenfratz and Niedermayer 1993), or 0.211 for a single free
fermion degree of freedom (Appendix). The result (3.31) matches neither
of these values.
This is not surprising, since the effective field theory at the critical point is expected to be a non-trivial
interacting theory. It might be possible to estimate this quantity via the $\epsilon$-expansion, using a
Landau-Ginzburg effective field theory. This has not yet been done, as far as we are aware.

\section{Conclusions}

	We have calculated the lowest-lying energy eigenvalues of the
transverse Ising model on the square lattice with periodic boundary
conditions for lattice sizes up to 6x6 sites, using the conjugate
gradient method. Finite-size scaling theory has been employed to
estimate the critical parameters, which are compared with previous
estimates in Table 5.

	It can be seen that our present estimates agree well with
earlier finite-size scaling results. We have achieved a substantial
increase in accuracy for the critical point, but only a more modest
increase for the critical index $\nu$. The results appear very
compatible with previous series analyses, and also with recent estimates
for the classical 3D Ising model, and field theory. This provides
further confirmation of the universality between these transitions.
Finally, it can be seen that the accuracy of the exponents for the
quantum model is now not very far behind that for the classical model.

	We have also estimated the finite-size scaling amplitudes for
the energy eigenvalues at the critical point. For the
spin gap we find
\begin{equation}
A_{1} = 1.39(1)
\end{equation}
to be compared with a previous estimate by Henkel (1987) of $A_{1} = 1.42$.

For the ground-state energy, we have shown evidence that
\begin{equation}
\epsilon_{0,L}(x_{L}) \sim \epsilon_{0}^{*} -
\frac{A_{0}}{L^{3}}, \hspace{5mm} L \rightarrow \infty
\label{4.2}
\end{equation}
and estimated
\begin{equation}
A_{0} = 0.35(2),
\end{equation}
to be compared with a previous estimate of 0.39 by Henkel (1987).
It should be possible to predict this amplitude from Landau-Ginzburg effective field theory.

	An extension to 7x7 sites of these exact diagonalization
calculations is hardly feasible at the present time, but there are some
very precise approximate methods now available, such as the density
matrix renormalization group (White 1992) and path integral Monte Carlo techniques (Sandvik 1992).
These might well be able to extend the results to larger
lattice sizes, and allow much improved finite-size scaling estimates of
the critical parameters. They could also confirm whether or not the
Casimir energy scales as in equation (\ref{4.2}). Our exact diagonalization
results should provide a useful calibration for such studies. We look forward to seeing such
calculations in the future.

	We have chosen here to work on the square lattice rather than
the triangular one, because the Hamiltonian matrix is somewhat smaller,
and the leading finite-size corrections are expected to be much the same
for both lattices. There are, however, some hints of irregularity or alternating
behaviour in some of the square lattice sequences. It might well be
that the triangular lattice results are smoother. 

\acknowledgements 

I would like to thank Dr. P.F. Price and Prof. I. Affleck for useful
discussions. Part of this work was carried out while on study leave at the
Institute for Theoretical Physics, University of California at Santa Barbara, and at
the Centre for Nonlinear Studies, Los Alamos National Laboratory. I
would like to thank Prof. R. Singh and the organizers of the Workshop
on `Magnetic Materials in Novel Materials and Geometries' for their
hospitality in Santa Barbara, and Dr. J. Gubernatis for his kind
hospitality in Los Alamos. The calculations were performed using
facilities at the New South Wales Centre for Parallel Computing, and at
the Centre for Nonlinear Studies, Los Alamos. I am very grateful to
Prof. R. Standish and Mr. D. Neal for their assistance in this regard.
This research was supported in part by the National Science Foundation
under grant no. PHY94-07194, and also by a grant from the Australian
Research Council.

\appendix{Appendix}

The finite-size scaling amplitude for the ground-state energy
(``Casimir amplitude") can be calculated for free fields as follows.

\subsection{ Free boson case}

	The zero-point energy of a free boson field is given in $d$ 
space dimensions by
\begin{equation}
E_{0} = \frac{1}{2}\sum_{k}\omega_{k}
\end{equation}
i.e. $\omega_{k}/2$ for each momentum mode. On a lattice, the free
particle Hamiltonian can be written in a finite-difference form
\begin{equation}
H = \frac{1}{2}\sum_{{\bf n}}[{\dot \phi}^{2}({\bf n}) + \sum_{i=1}^{d}
(\frac{\phi({\bf n+i}) - \phi({\bf n-i})}{2})^{2}]
\end{equation}
where the lattice spacing has been set to 1. The eigenmodes are plane waves
\begin{equation}
\phi({\bf n},t) =\frac{1}{N^{1/2}}\sum_{{\bf k}}[a_{k}e^{i({\bf
k.n}-\omega_{k}t)} + a_{k}^{\dagger}e^{-i({\bf k.n}-\omega_{k}t)}]
\end{equation}
whence
\begin{equation}
\omega_{k} = [2\sum_{i=1}^{d}(1-\cos k_{i})]^{1/2}
\end{equation}
and for periodic boundary conditions the allowed momenta are
\begin{equation}
k_{i} = \frac{2\pi}{L}l_{i}, \hspace{5mm} l_{i} = 0,1,2,\cdots
\end{equation}
for a lattice of $N=L^{d}$ sites. Hence
\begin{equation}
E_{0} = \frac{1}{2}\sum_{{\bf k}}[2\sum_{i=1}^{d}(1-\cos k_{i})]^{1/2}
\end{equation}
i.e.
\begin{equation}
\epsilon_{0} = \frac{1}{L^{d}}\sum_{\{l_{i}\}=0}^{L-1}[\sum_{i=1}^{d}\sin^{2}(\frac{\pi
l_{i}}{L})]^{1/2}
\end{equation}
Now the leading finite-size correction to this sum arises from the
infrared (small momentum) behaviour of the lattice sum 
(Hasenfratz and Leutwyler 1990), and does not depend on the cutoff or regularization at
large momentum. Thus we may approximate for our purposes
\begin{equation}
\epsilon_{0} \simeq \frac{1}{L^{d}}\sum_{\{l_{i}\}= -\infty}^{\infty}[\sum_{i=1}^{d}(\frac{\pi
l_{i}}{L})^{2}]^{1/2}
\end{equation}
Now use the Poisson resummation formula
\begin{equation}
\sum_{{\bf m}=-\infty}^{+\infty}f({\bf m}L) = \frac{1}{L^{d}}\sum_{{\bf
n}=-\infty}^{+\infty}g(\frac{2\pi{\bf n}}{L})
\end{equation}
with
\begin{equation}
g({\bf k}) = \int_{-\infty}^{+\infty}e^{i{\bf k.x}}f({\bf x})d^{d}x
\end{equation}
to show
\begin{equation}
\epsilon'_{0}
=\frac{1}{4\pi}\sum_{\{m_{i}\}'=-\infty}^{+\infty}\int_{0}^{\infty}k^{d}dk
\frac{J_{d/2-1}(kx)}{(2\pi kx)^{d/2-1}}\hspace{5mm} (x = L|{\bf m}|)
\end{equation}

\begin{equation}
= - \frac{\Gamma(\frac{d+1}{2})}{2\pi^{\frac{d+1}{2}}L^{d+1}}\sum_{\{m_{i}\}'
= -\infty}^{+\infty}\frac{1}{|{\bf m}|^{d+1}}
\end{equation}

The dash here implies removal 
of the term ${\bf m}=0$, which corresponds to the (infinite,
non-universal) {\it bulk} ground-state energy per site, which we simply
drop. 

	The sum involved here is a generalization of the Riemann zeta
function. It gives
\begin{equation}
\epsilon_{0}' = -\frac{A_{0}}{L^{d+1}}
\end{equation}
where for $d = 1$, $A_{0}$ is easily evaluated
\begin{equation}
A_{0} = \frac{\pi}{6} = 0.5236
\end{equation}
the result familiar from conformal field theory. For higher
dimensions, we have evaluated the sum numerically
\begin{equation}
d = 2: A_{0} =  0.7189
\end{equation}
\begin{equation}
d = 3: A_{0} =  0.8375
\end{equation}
The result for $d=2$ was given previously by Hasenfratz and Niedermayer (1993)
.

\subsection{Free fermion case}

A similar naive argument can be given for the case of a single species
of free Weyl (spinless) fermions. The filled Dirac sea has
energy
\begin{equation}
E_{0} = -\sum_{k}\omega_{k}
\end{equation}
where again
\begin{equation}
\omega_{k} = [2\sum_{i=1}^{d}(1-\cos k_{i})]^{1/2}
\end{equation}
and we assume antiperiodic boundary conditions for the fermions
\begin{equation}
k_{i} = \frac{\pi}{L}(2l_{i}+1), \hspace{5mm}l_{i} = 0,1,2,\cdots
\end{equation}
Hence
\begin{equation}
\epsilon_{0} \simeq -\frac{1}{L^{d}}\sum_{\{l_{i}\}= -\infty}^{\infty}[\sum_{i=1}^{d}(\frac{\pi
(2l_{i}+1)}{2L})^{2}]^{1/2}
\end{equation}
Use the Poisson resummation formula again to find
\begin{equation}
\epsilon'_{0}
=-\frac{1}{4\pi}\sum_{\{m_{i}\}'=-\infty}^{+\infty}(-1)^{\sum_{i}m_{i}}\int_{0}^{\infty}k^{d}dk
\frac{J_{d/2-1}(kx)}{(2\pi kx)^{d/2-1}} \hspace{5mm}(x = L|{\bf m}|)
\end{equation}

\begin{equation}
=  \frac{\Gamma(\frac{d+1}{2})}{2\pi^{\frac{d+1}{2}}L^{d+1}}\sum_{\{m_{i}\}
'=-\infty}^{+\infty}\frac{(-1)^{\sum_{i}m_{i}}}{|{\bf m}|^{d+1}}
\end{equation}

For $d=1$ $A_{0}$ is easily evaluated to give

\begin{equation}
A_{0} = \frac{\pi}{12} = 0.2618
\end{equation}

also familiar from conformal field theory; while for higher
dimensions, we find numerically

\begin{equation}
d = 2: A_{0} =  0.2106
\end{equation}

\begin{equation}
d = 3: A_{0} =  0.1957
\end{equation}
These numbers have not been obtained previously, as far as we are aware.

\figure{{\bf Figure 1.} 
Graph of the finite-lattice pseudo-critical points $x_{L}$ as a function of $1/L^{4}$. The line is merely to guide the
eye.
\label{fig1}}
\figure{{\bf Figure 2.} 
Finite-lattice estimates of the index $1/\nu$ graphed against $1/L$. The full circles are `linear' estimates, the open
circles are `logarithmic' estimates. The lines are merely to guide the eye.
\label{fig2}}
\figure{{\bf Figure 3.} 
The energy gap amplitude $A_{1,L}$ graphed against $1/L^{2}$.
\label{fig3}}
\figure{{\bf Figure 4.} 
The ground-state energy per site at lattice size $L$ graphed against $1/L^{3}$.
\label{fig4}}
\figure{{\bf Figure 5.} 
The effective exponent $p_{L}$ graphed against $1/L$.
\label{fig5}}
\figure{{\bf Figure 6.} 
The Casimir amplitude $A_{0,L}$ graphed against $1/L$.
\label{fig6}}
\figure{{\bf Figure 7.} 
Finite-lattice estimates of the `speed of light' $v$ graphed against $1/L$.
\label{fig7}}

\newpage
\begin{table}
\caption{
Finite-lattice data at the pseudo-critical points $x_{L}$, calculated
for the pair of lattice sites L and (L-1) in each case. Given are the
ground-state energy per site $\epsilon_{0}$, its first two derivatives
$\epsilon'_{0}$ and $\epsilon''_{0}$ with respect to x, and the
susceptibility $\chi$.}
\begin{tabular}{cccccc}  
\multicolumn{1}{c}{$x_{L}$}  & \multicolumn{1}{c}{$L$} & 
\multicolumn{1}{c}{ $\epsilon_{0}$} & 
\multicolumn{1}{c}{ $\epsilon_{0}'$} & 
\multicolumn{1}{c}{ $\epsilon_{0}''$} & 
\multicolumn{1}{c}{ $\chi$}  \\ \tableline 
 0.26034238222 &  1 & -0.520684764436 & -2.00000000 &  0.00000
& -0.09000 \\
               &  2 & -0.074060535180 & -0.61163273 & -2.90881
& -0.32656 \\
 0.31600008772 &  2 & -0.112709188696 & -0.77866508 & -3.06919
& -0.453734 \\
               &  3 & -0.070818161984 & -0.59879739 & -4.15119 & -1.00929 \\
 0.32424925229 &  3 & -0.075901188836 & -0.63384113 & -4.34370 & -1.13177 \\
               &  4 & -0.066430308096 & -0.57114518 & -5.09931
& -1.99844 \\
 0.32669593806 &  4 & -0.067843120636 & -0.58378935 & -5.23641
& -2.11883 \\
               &  5 & -0.064637823298 & -0.55295527 & -5.83324
& -3.29209 \\
 0.32758326752 &  5 & -0.065130784765 & -0.55817035 & -5.92135 &  - \\
               &  6 & -0.063752757694 & -0.54012509 & -6.42027&
- 
\end{tabular}
\end{table}

\begin{table}
\caption{
Finite-lattice data as in Table 1, consisting of the mass gap F, its
first two derivatives$ F'$ and$ F''$ with respect to x, and the
`magnetization' M.}
\begin{tabular}{cccccc} 
 \multicolumn{1}{c}{$x_{L}$}  & \multicolumn{1}{c}{ $L$}  & 
\multicolumn{1}{c}{ $F$}  & 
\multicolumn{1}{c}{ $F'$}  & 
\multicolumn{1}{c}{ $F''$}  & 
\multicolumn{1}{c}{ $M$}  \\ \tableline 
 0.26034238222 &  1 & 2.000000000000 &  0.0000000 & 0.00000 & 1.00000 \\
               &  2 & 1.000000000000 & -3.4007921 & 6.05294 & 0.67337 \\
 0.31600008772 &  2 & 0.820891162135 & -3.0223297 & 7.44433 & 0.71908 \\
               &  3 & 0.547260774756 & -4.4695972 & 13.9564 & 0.58357 \\
 0.32424925229 &  3 & 0.510885529302 & -4.3470817 & 15.7416 & 0.59705 \\
               &  4 & 0.383164146983 & -5.4088281 & 25.0603 & 0.51520 \\
 0.32669593806 &  4 & 0.370007329425 & -5.3452110 & 26.9450 & 0.52130 \\
               &  5 & 0.296005863592 & -6.2337554 & 39.3896 & 0.46488 \\
 0.32758326752 &  5 & 0.290490201648 & -6.1980387 & 41.1167 &
- \\
               &  6 & 0.2420751678 & -6.9850090 & 56.7636 &
-
\end{tabular}
\end{table}
\newpage

\begin{table}
\caption{
Finite-size scaling estimates of the critical indices, as defined by
equation (\ref{3.7}) and following, where L is the larger of the two
lattice sizes used in the estimate.
}\label{tab3}  
\begin{tabular}{cccccc} 
\multicolumn{1}{c}{ L } &\multicolumn{1}{c}{ $1/\nu$}
  &\multicolumn{1}{c}{ $1 - \alpha/\nu$} 
&\multicolumn{1}{c}{ $\beta/\nu$ }
&\multicolumn{1}{c}{ $\gamma/\nu$ }
&\multicolumn{1}{c}{ $\rho$ }
  \\ \tableline 
   2 & 1.27817 & -  & 0.653264 & 1.44880 & - \\
   3 & 1.38021 & 1.88408  & 0.565352 & 1.65132 & 1.71862 \\
   4 & 1.43242 & 1.20661  & 0.548326 & 1.73469 & 2.09567 \\
   5 & 1.46377 & 1.05069  & 0.541083 & 1.78195 & 2.30781 \\
   6 & 1.48479 & 0.98436  & -        & -       & 2.42047 
 \\ \tableline
$\infty$ & 1.591(2) & 0.84(1) & 0.523(2) & 1.95(1) & 2.8(2)
\end{tabular}
\end{table}

\begin{table}
\caption{ 
`Logarithmic' finite-size scaling estimates of the critical indices, as defined by
equation (\ref{3.11}) and following, where L is the larger of the two
lattice sizes used in the estimate.
}\label{tab4} 
\begin{tabular}{cccccc} 
\multicolumn{1}{c}{  L}
 & \multicolumn{1}{c}{  $1/\nu$}
 & \multicolumn{1}{c}{  $1 - \alpha/\nu$}
 & \multicolumn{1}{c}{  $\beta/\nu$}
 & \multicolumn{1}{c}{  $\gamma/\nu$}
 & \multicolumn{1}{c}{  $\rho$}
  \\ \tableline 
   2&  - & - & - & - & - \\
   3&  1.52002 & 2.43901  &  0.514989 & 1.97178 &  4.83686709 \\
   4&  1.54105 & 1.24804  &  0.512493 & 1.97642 &  4.16837815 \\
   5&  1.55226 & 1.05715  &  0.513266 & 1.97479 &  3.85524695 \\
   6&  1.55937 & 0.98287  &  -        & -       &  3.63001774 
 \\ \tableline
$\infty$ & 1.593(3) & 0.84(1) & 0.521(3) & 1.96(1) & 2.5(2)
\end{tabular}
\end{table}

\begin{table}
\caption{ 
A comparison of critical parameters obtained in the present work with some others obtained
elsewhere. Key: HT = high-temperature series; LT = low-temperature series; FS = finite-size
scaling; MC = Monte Carlo; TR = triangular lattice; SQ = square lattice
}\label{tab5} 
\begin{tabular}{llllll} 
\multicolumn{1}{c}{  }
 & \multicolumn{1}{c}{  $\nu$}
 & \multicolumn{1}{c}{  $\alpha$}
 & \multicolumn{1}{c}{  $\beta$}
 & \multicolumn{1}{c}{  $\gamma$}
 & \multicolumn{1}{c}{  $x_{c}$}
  \\ \tableline 
{\it (2+1)-dimensional Ising model}& & & & &  \\
FS SQ Hamer (1983) & 0.635(5) & & & & 0.3289(10) \\
FS SQ Henkel (1984, 1987) & 0.629(2) & 0.11(1) & 0.324(9) & - & 0.3282(1) \\
FS SQ This work & 0.629(1) & 0.10(1)  & 0.328(2) & 1.23(1) & 0.32841(2) \\
FS TR Hamer \& Johnson (1986) & 0.627(4) & - & 0.332(6) & 1.236(8) & \\
FS TR Price et al (1993) & 0.627(2) & 0.12(2) & 0.324(3) & 1.23(1) & \\
HT SQ He et al (1990) & 0.637(4) & 0.11(2) & & 1.244(4) & 0.32851(8) \\
LT SQ Oitmaa et al (1991) & 0.64(3) & 0.096(6) & 0.318(4) & 1.25(2) & \\
{\it 3-dimensional Ising model} & & & & & \\
HT BCC Butera and Comi (1997) & 0.6308(5) & & & 1.2384(6) & \\
MC Hasenbusch (1999) & 0.6296(3)(4) & & & 1.2367(11) & \\
{\it Field Theory} & & & & & \\
Guida and Zinn-Justin (1997) & 0.6304(13) & 0.109(4) & 0.3258(14) & 1.2396(13) 
\end{tabular}
\end{table}


\begin{references}
\bibitem[*]{cjh}Email address: c.hamer@unsw.edu.au

\bibitem{} Barber M N 1983 {\it Phase Transitions and Critical
Phenomena} vol 8, ed C Domb and J Lebowitz (New York: Academic)
\bibitem{} Barber M N and Hamer C J 1982 J. Aust. Math. Soc. {\bf B23} 229
\bibitem{} Bulirsch R and Stoer J 1964 Numer. Math. {\bf 6} 413 
\bibitem{} Butera P and Comi M 1997 Phys. Rev. {\bf B56} 8212
\bibitem{} Cardy J L 1985 J. Phys. A: Math. Gen. {\bf 18} L757
\bibitem{} Fradkin E and Susskind L 1978 Phys. Rev. {\bf D17} 2637
\bibitem{} Guida R and Zinn-Justin J 1998 J. Phys. {\bf A31} 8103
\bibitem{} Guttman A J 1989 {\it Phase Transitions and Critical
Phenomena} vol 13, ed C Domb and J Lebowitz (New York: Academic)
\bibitem{} Hamer C J 1982 J. Phys. A: Math. Gen. {\bf 15} L675
\bibitem{} Hamer C J 1983 J. Phys. A: Math. Gen. {\bf 16} 1257
\bibitem{} Hamer C J and Guttmann A J 1989 J. Phys. A: Math. Gen. {\bf 22} 3653
\bibitem{} Hamer C J and Johnson C H J 1986 J. Phys. A: Math. Gen. {\bf 19}
423 
\bibitem{} Hamer C J and Irving A C 1984 J.Phys. A: Math. Gen. {\bf 17} 1649 
\bibitem{} Hasenbusch M 1999 J. Phys. A: Math. Gen. {\bf 32} 4851
\bibitem{} Hasenfratz P and Leutwyler H 1990 Nucl. Phys. {\bf B343} 241
\bibitem{has93}Hasenfratz P and Niedermayer F 1993 Z. Phys. {\bf B92}, 91 
\bibitem{} He H-X, Hamer C J and Oitmaa J 1990 J Phys A: Math. Gen. {\bf
23}
1775
\bibitem{} Henkel M 1984 J. Phys. A: Math. Gen. {\bf 17} L795 
\bibitem{} Henkel M 1986 J. Phys. A: Math. Gen. {\bf 19} L247
\bibitem{} Henkel M 1987 J. Phys. A: Math. Gen. {\bf 20} 3969
\bibitem{} Henkel M 1990 in {\it Finite-Size Scaling and Numerical
Simulation of Statistical Systems} ed V Privman (Singapore: World
Scientific)
\bibitem{} Henkel M and Patkos A 1987 J. Phys. A: Math. Gen. {\bf 20} 2199
\bibitem{} Henkel M and Sch{\" u}tz G 1988 J. Phys. A: Math. Gen. {\bf 21}
2617 
\bibitem{lin90}Lin H Q 1990 Phys. Rev. {\bf B42}, 6561 
\bibitem{} Lubkin S 1952 J. Res. NBS {\bf 48} 228
\bibitem{} Marland L G 1981 J Phys A: Math. Gen. {\bf 14} 2047
\bibitem{} Mon KK 1985 Phys. Rev. Lett. {\bf 54} 2671
\bibitem{nig93}Nightingale M P, Visvanath V S and M{\" u}ller G 1993 Phys. Rev. {\bf B48}, 7696
\bibitem{} Oitmaa J, Hamer C J and Zheng W-H 1991 J. Phys. A: Math. Gen. {\bf 24} 2863
\bibitem{} Pfeuty P and Elliott R J 1971 J. Phys. C: Solid State Phys. {\bf 4}
2370
\bibitem{} Price P F, Hamer C J and O'Shaughnessy D 1993 J. Phys. A:
Math. Gen. {\bf 26} 2855
\bibitem{} Privman V and Fisher M E 1984 Phys. Rev. {\bf B30} 322
\bibitem{} Roomany H and Wyld H W 1980 Phys. Rev. {\bf D21} 3341
\bibitem{} Sandvik A W 1992 J. Phys. {\bf A25} 3667
\bibitem{} Schulz H J, Ziman T A L and Poilblanc D 1996 J. Physique {\bf
I6} 675
\bibitem{} Smith D A and Ford W F 1982 Math. Comp. {\bf 38} 481
\bibitem{} Suzuki M 1976 Prog. Theor. Phys. {\bf 56} 1454
\bibitem{} Uzelac K 1980 Thesis Orsay
\bibitem{} van den Broeck J-M and Schwartz L W 1979 SIAM J. Math. Anal. {\bf 10} 658
\bibitem{} Weigel M and Janke W 1999 Phys. Rev. Letts. {\bf 82} 2318
\bibitem{} Weston R A 1990 Phys. Lett. {\bf B248} 340
\bibitem{} White S R 1992 Phys. Rev. Letts. {\bf 69} 2863
\bibitem{ } Yanase A, Takeshige Y and Suzuki M 1976 J. Phys. Soc. Japan
{\bf 41} 1108
\bibitem{} Yang C N 1952 Phys. Rev. {\bf 85} 808
\end{references}
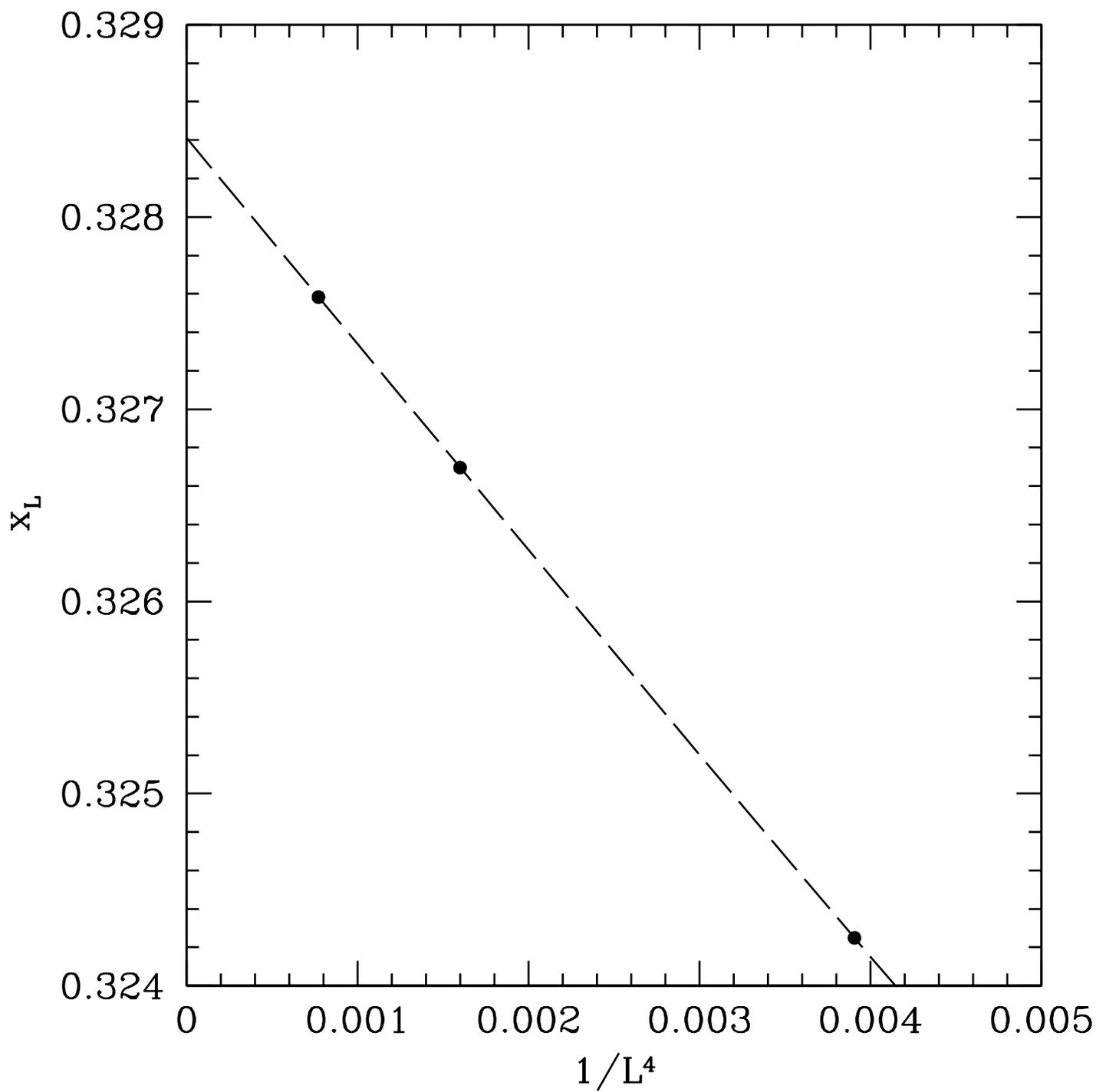
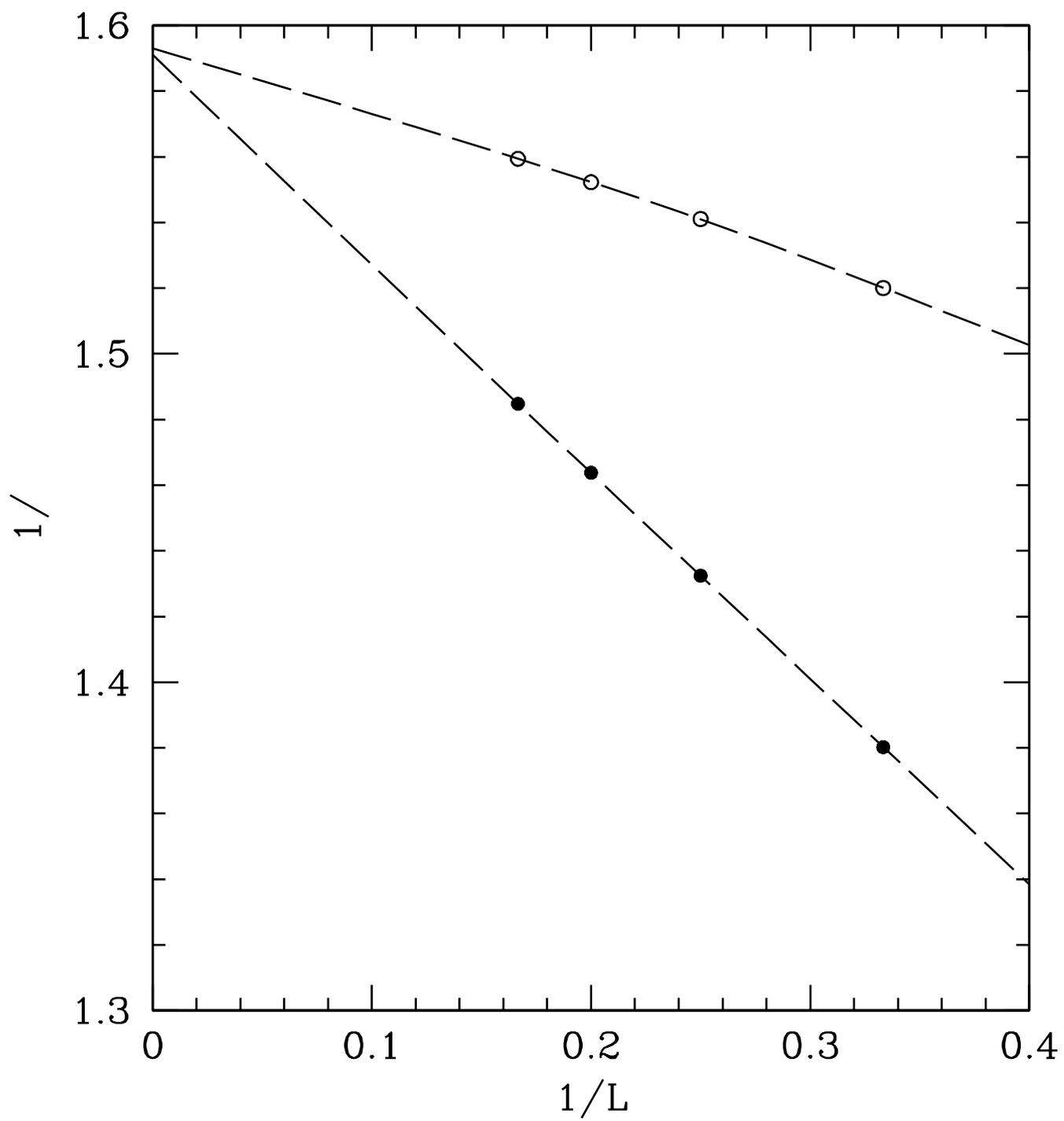
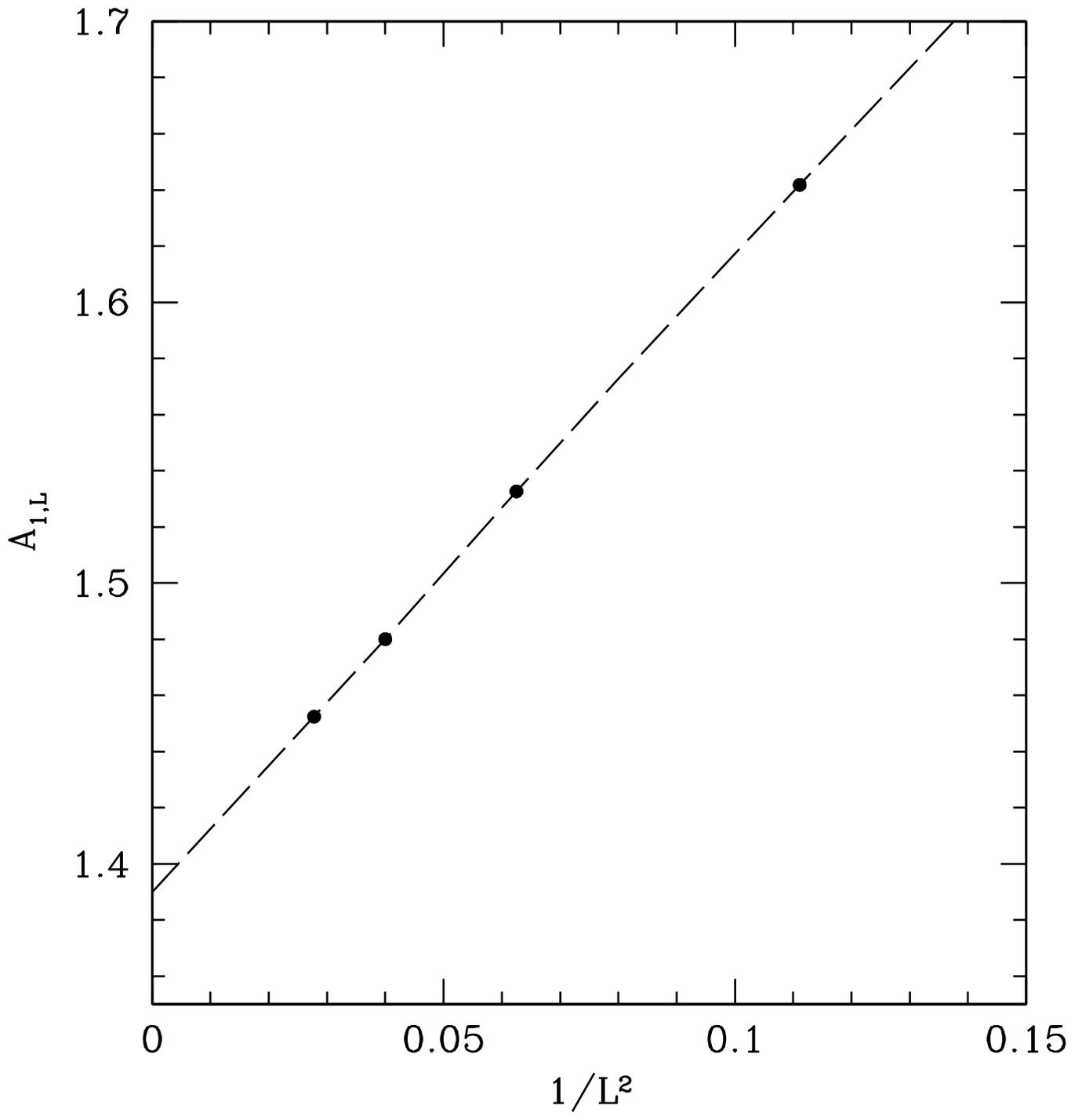
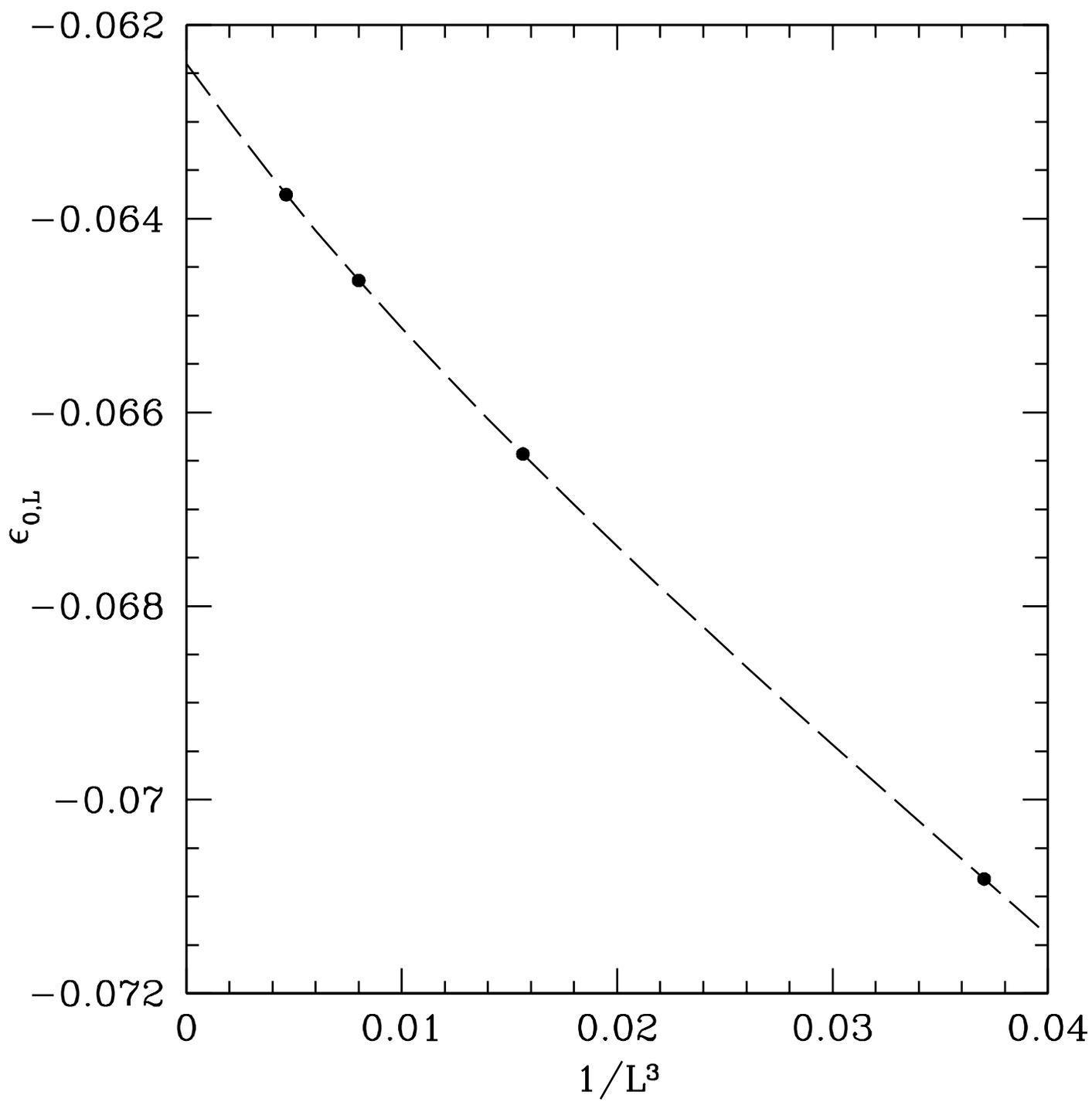
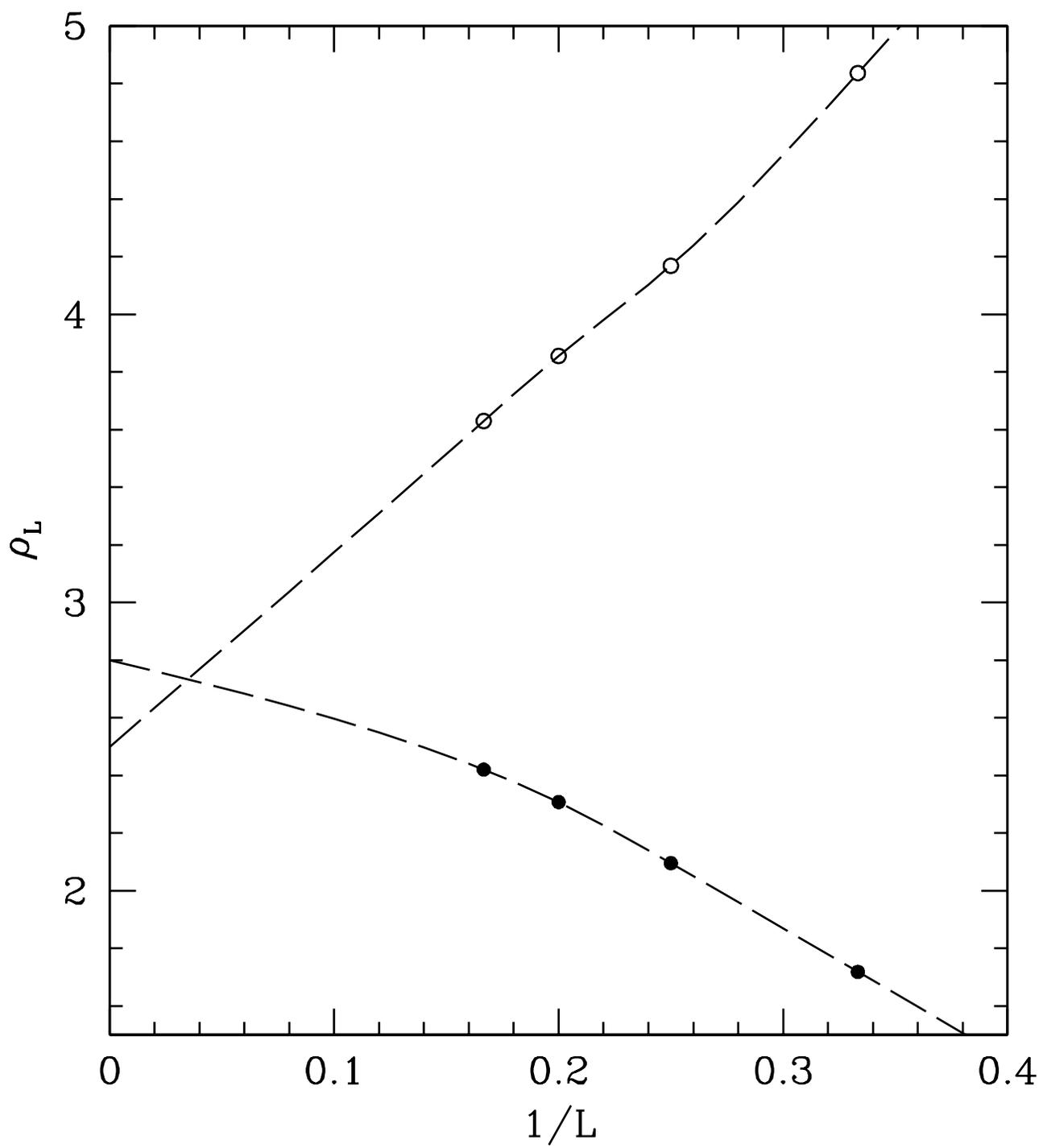
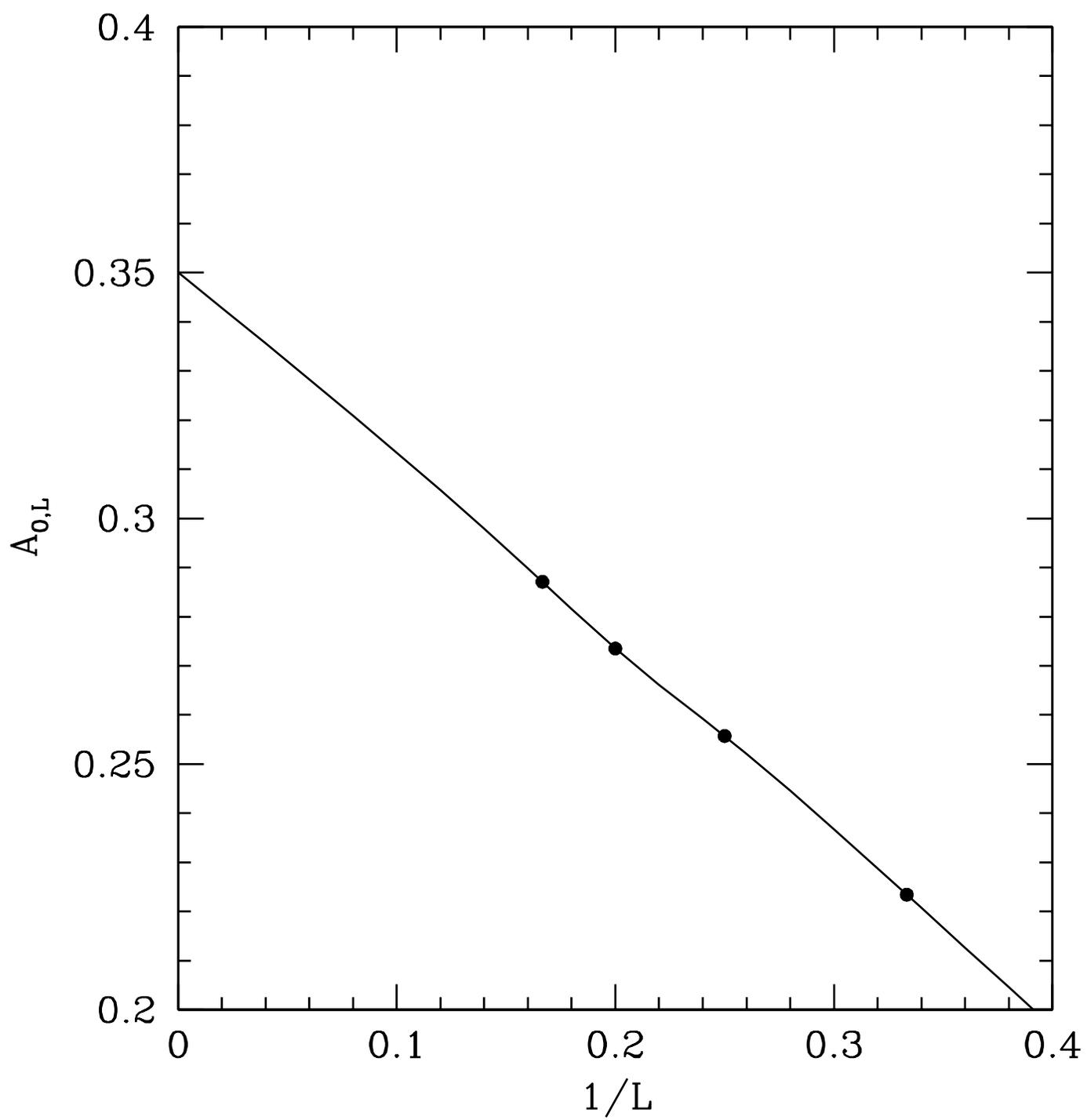
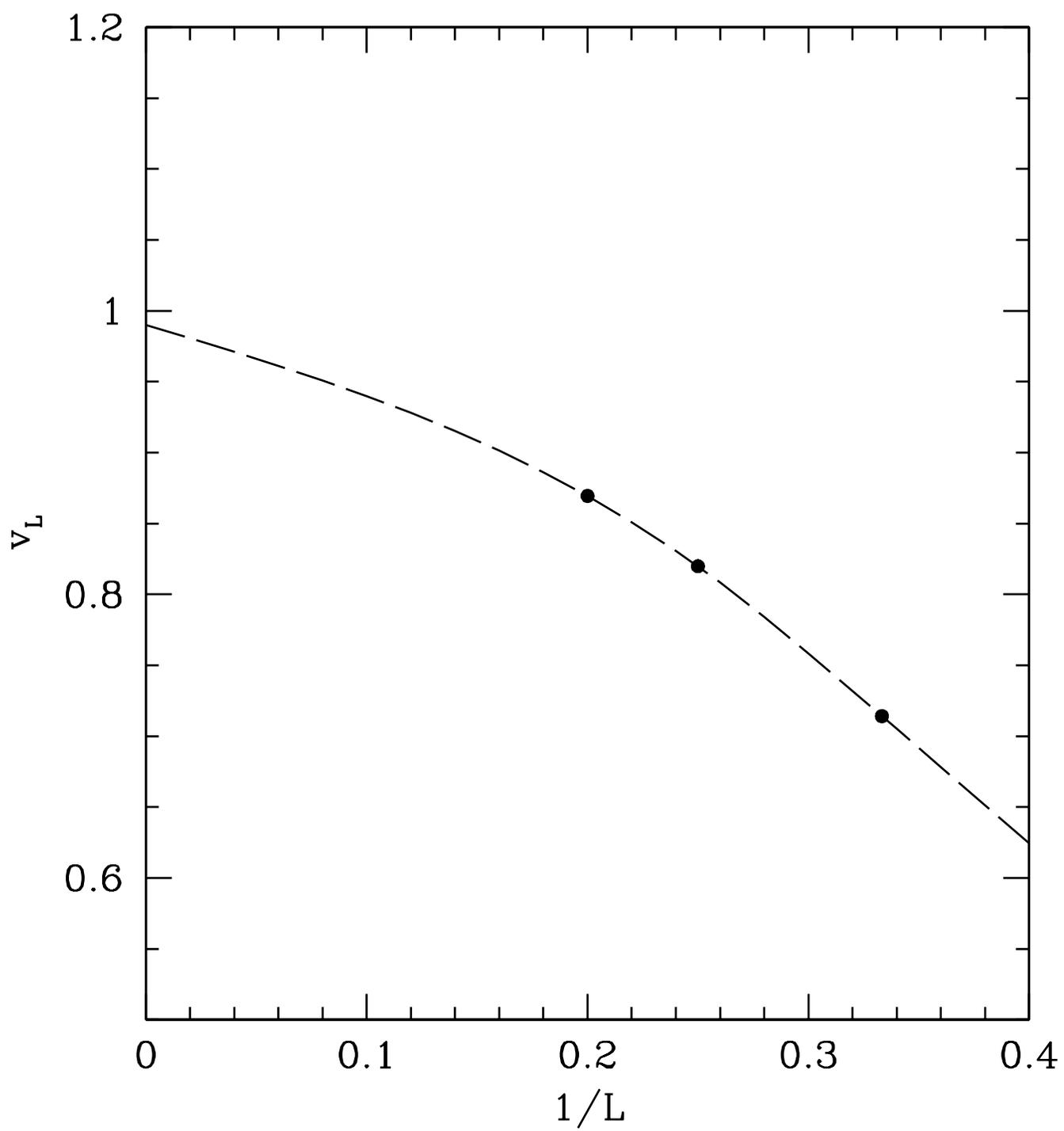
\end{document}